# The Discrete Hilbert Transform for Non-Periodic Signals

Sumanth Kumar Reddy Gangasani

## Abstract

This note investigates the size of the guard band for non-periodic discrete Hilbert transform, which has recently been proposed for data hiding and security applications. It is shown that a guard band equal to the duration of the message is sufficient for a variety of analog signals and is, therefore, likely to be adequate for discrete or digital data.

## Introduction

The discrete Hilbert transform (DHT) has a variety of applications in signal representation and processing [2-4]. The transform comes in a variety of forms based on assumptions on how the signal is derived from a corresponding periodic signal. The form presented by Kak [1] takes the signal to be non-periodic. Several versions of the transform for periodic signals are also known [5], [6], [7], [8], and these have been extensively examined in the literature.

Here we consider the non-periodic discrete Hilbert transform, which has not received as much of attention as it deserves, and investigate the question of the size of the guard band that is optimal for its use. Further motivation for this study is that Hilbert transformation has recently been used for data hiding [9]. It can also be employed for waveform scrambling [10], [11] as in the new invention on masked audio signals [12].

The guard band could, in principle, be used to hide additional information. Therefore, its use should not be considered as an additional burden that has no redeeming value.

## The Problem

The problem we consider in this paper is that of approximation of DHT using a finite number of data points. Specifically, let *f(n)* be non-zero over (0, *n*-1), then we wish to minimize *m* so that

$$\sqrt{\frac{\sum_{m=0}^{k}\left(f(n)-f'_m(n)\right)^2}{n}} < \theta$$

where

$f'_m(n)$ = $\text{DHT}^{-1}$(g(k)) over (*-m, n+m-1)* and $\theta$ is the threshold value for the error.

In other words, we are considering a guard band of *m* points on both sides of the signal. The restrictions on the value of $\theta$ for a continuous message would be stringent, but for discrete data (the kind used in cryptography applications) these would be much less stringent, because the quantization inherent in the data would allow its reconstruction even in the case where there is some residual RMS error between the original signal and its DHT transform.

## Definition of DHT

In this section begin with the DHT formula in reference [1].

$$DHT\{f(n)\} = g(k) = \begin{cases} \dfrac{2}{\pi} \displaystyle\sum_{n=odd} \dfrac{f(n)}{k-n}; & k \ even \\ \dfrac{2}{\pi} \displaystyle\sum_{n=even} \dfrac{f(n)}{k-n}; & k \ odd \end{cases}$$

Inverse DHT is given by

$$f(n) = \begin{cases} \dfrac{-2}{\pi} \displaystyle\sum_{k=odd} \dfrac{g(k)}{n-k}; & n \ even \\ \dfrac{-2}{\pi} \displaystyle\sum_{k=even} \dfrac{g(k)}{n-k}; & n \ odd \end{cases}$$

Since DHT is defined over all positive and negative integers, one needs to investigate how many extra points are required to be taken into consideration in the DHT domain.

Since the denominator is linear, the effect of the message points would tend to spread out in the transform domain.

## Results

We have performed experiments on a variety of analog signals to find the relationship between RMS error and the size of the guard band. The consideration of analog signals was prompted by the fact that this helps one find the constraints much better than consideration of digital signals.

Sine, ramp, square, and triangle signals were considered for the experiments. The signal width was 90 points and we considered guard bands ranging from 0 to 900.

In each of the four cases considered, the RMS error became quite close to zero as the guard band approached 90. Expectedly, the worst result was obtained for the square signal because for this case the value of the signal points at the edges of the signal range is high and, therefore, its effect is spread out further into the guard band.

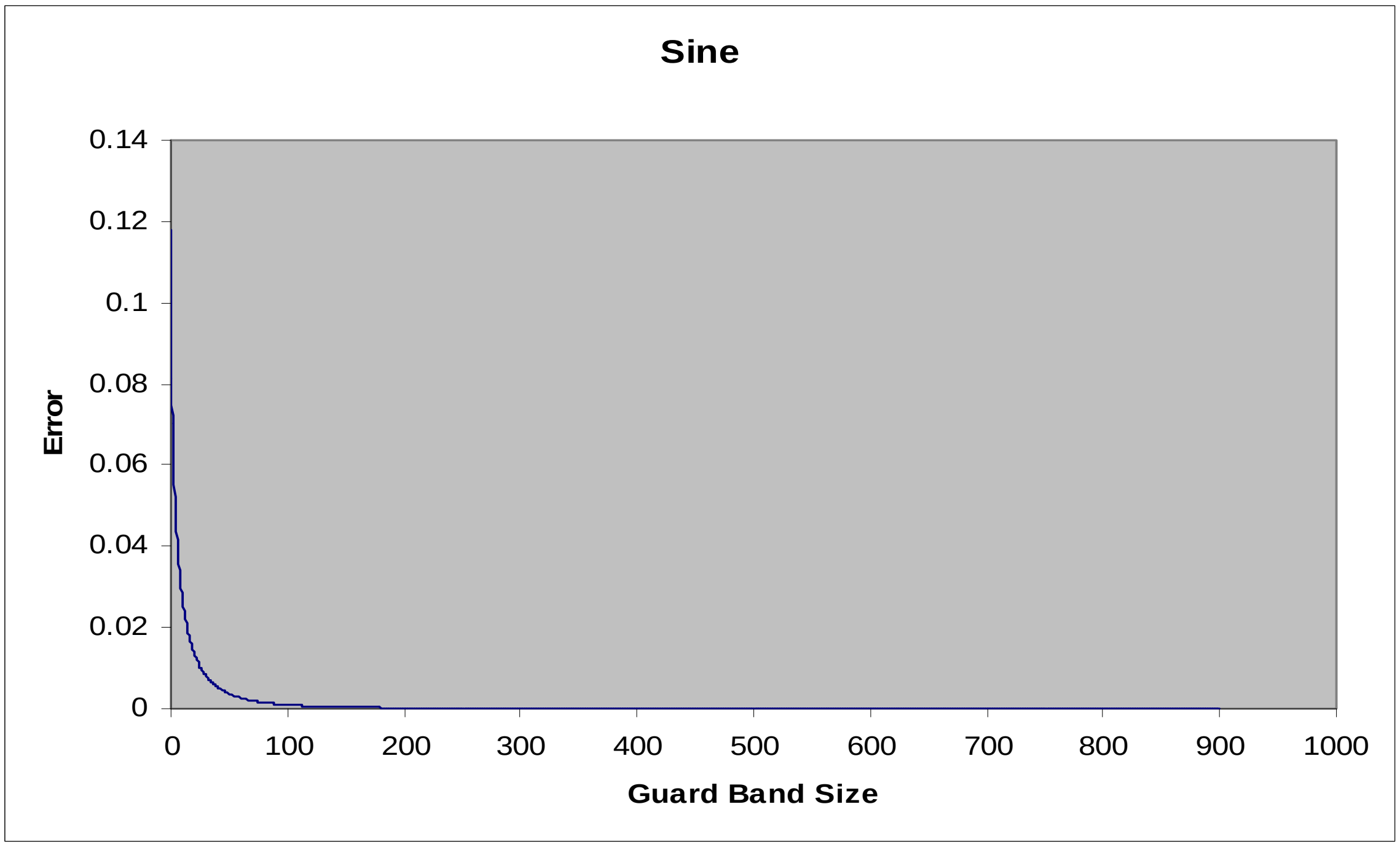

For the sine signal, the value of the RMS error for guard band equal to 90 is 1.02% of RMS error without guard band.

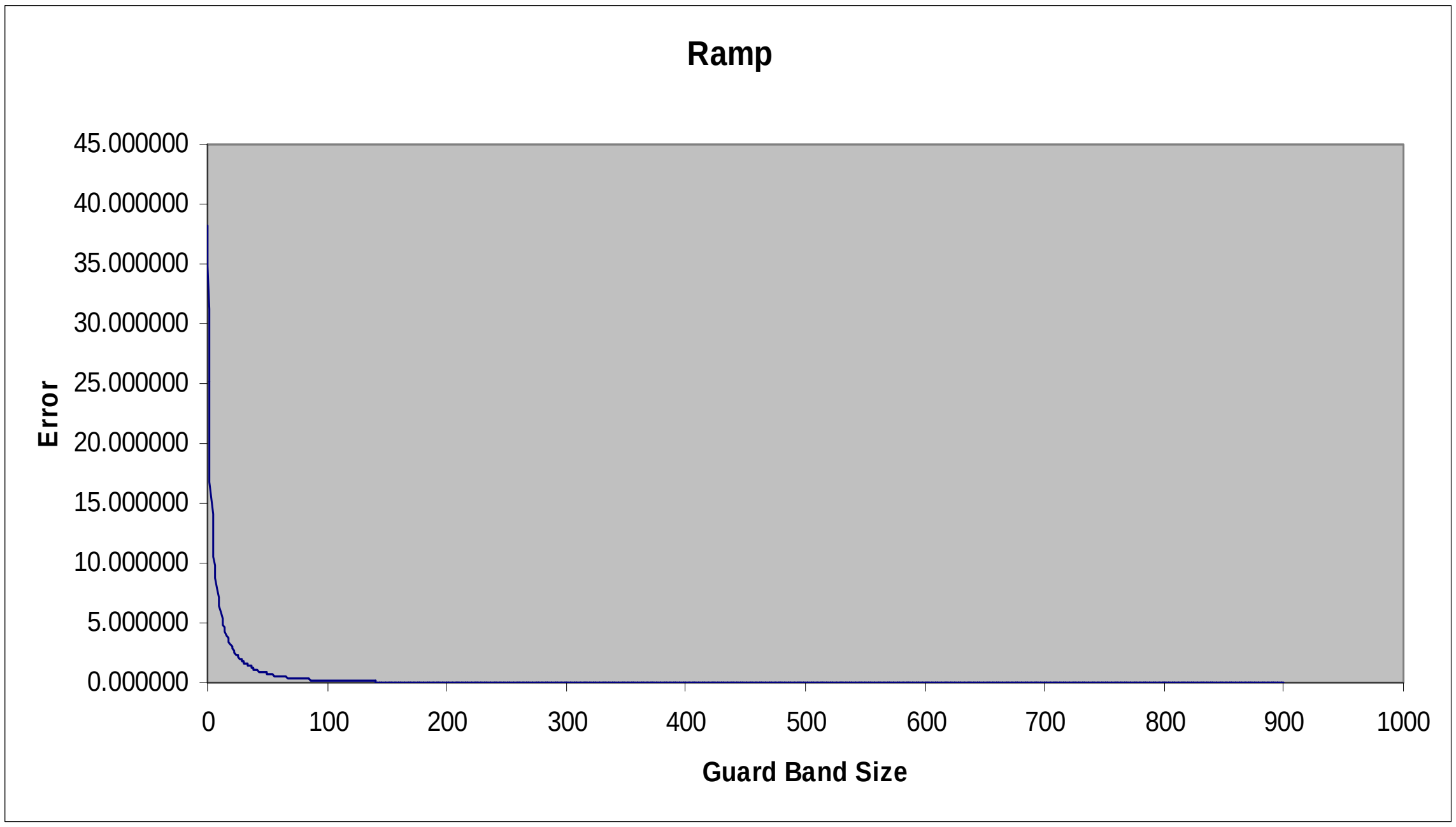

For the ramp signal, the value of the RMS error for guard band equal to 90 is 0.62% of RMS without guard band.

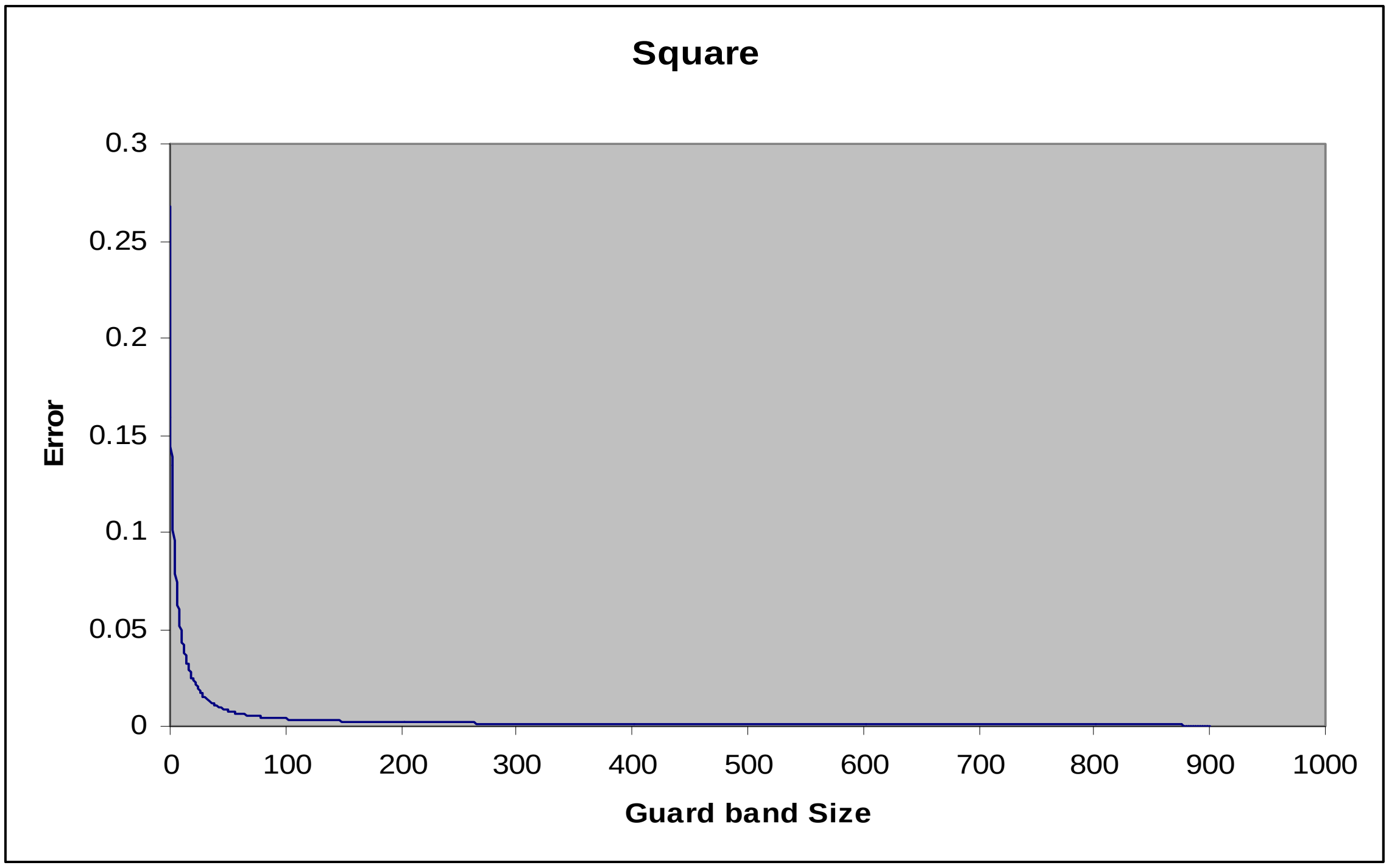

For the square signal, the value of the RMS error for guard band equal to 90 is 1.6% of RMS without guard band.

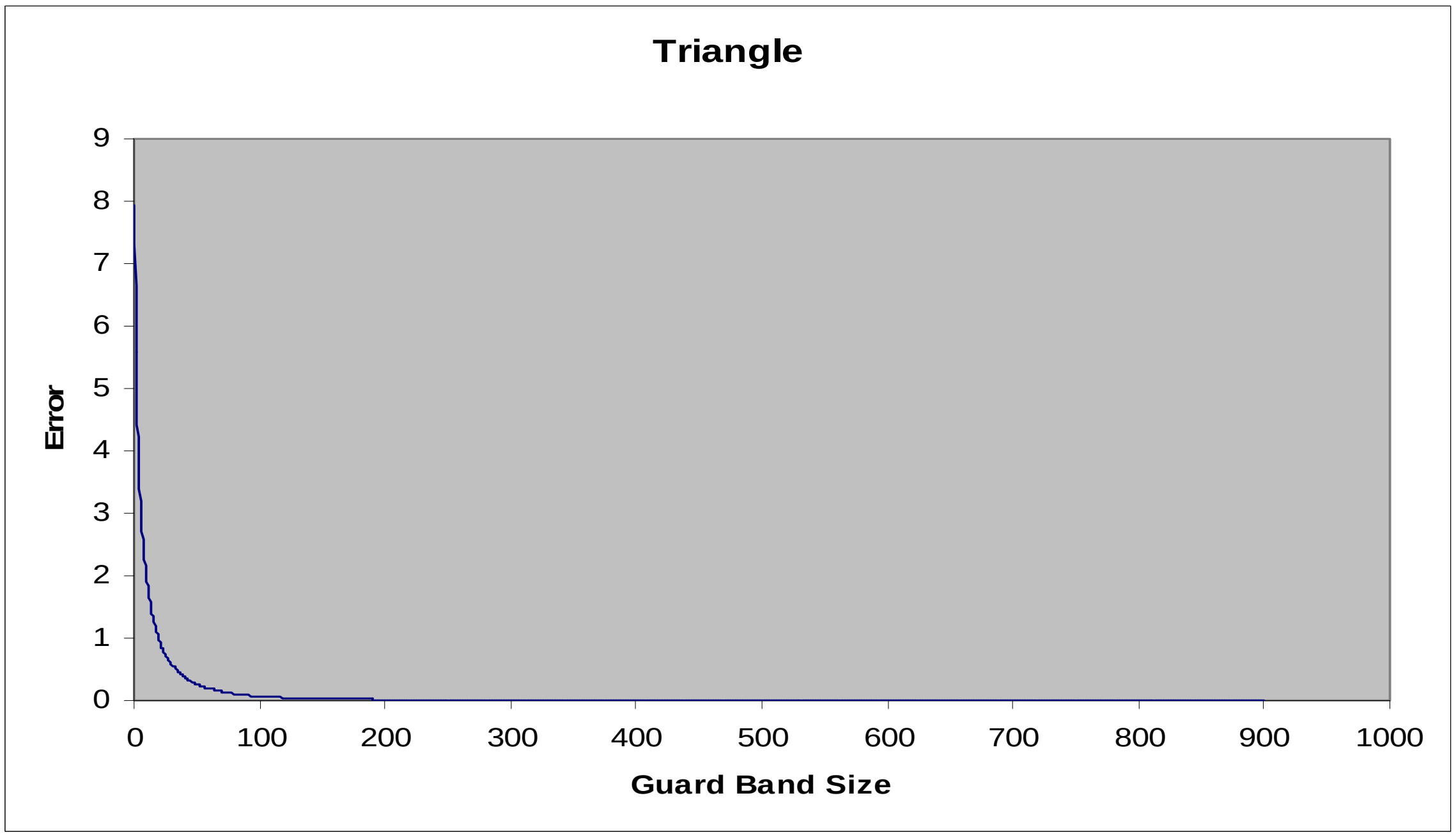

For the triangle signal, the value of the RMS error for guard band equal to 90 is 1.08% of RMS without guard band.

In each of these cases, the RMS error for guard band equal to the message width is much less than 2%.

## Conclusions

We have shown that the non-periodic discrete Hilbert transform can be effectively used for finite sequences. We have performed experiments and determined that to use a guard band equal to the width of the signal that is transform domain having a total of 3 times as many points as in the signal domain, works very well.

The guard band can be used for hiding of information. This opens up the possibility of using DHT for many date security applications.